\documentclass{ws-ijmpd}
\usepackage[super,compress]{cite}
\begin{document}

\markboth{Ramon Lapiedra and Juan Antonio Morales-Lladosa}
{Spherical Symmetric parabolic dust collapse: ${\cal C}^{1}$ matching metric with zero intrinsic energy}

\catchline{}{}{}{}{}
%
%%%%%%%%%%%%%%%%%%%%%%%%%%%%%%%%%%%%%%%%%%%%%%%%%%%%%%%%%%%%%%%%%%%%

\title{SPHERICAL SYMMETRIC PARABOLIC DUST COLLAPSE: \\ ${\cal C}^{1}$ MATCHING METRIC WITH ZERO INTRINSIC ENERGY}

\author{RAMON LAPIEDRA}

\address{Departament d'Astronomia i Astrof\'{\i}sica, Universitat de
Val\`encia, \\E-46100 Burjassot, Val\`encia, Spain.\\
Observatori Astron\`omic, Universitat de
Val\`encia, E-46980 Paterna, Val\`encia, Spain.\\
ramon.lapiedra@uv.es}

\author{JUAN ANTONIO MORALES--LLADOSA}
\address{Departament d'Astronomia i Astrof\'{\i}sica, Universitat de
Val\`encia, \\E-46100 Burjassot, Val\`encia, Spain.\\
Observatori Astron\`omic, Universitat de
Val\`encia, E-46980 Paterna, Val\`encia, Spain.\\
antonio.morales@uv.es}

\maketitle{}

%\begin{history}
%\received{Day Month Year}
%\revised{Day Month Year}
%\end{history}

\begin{abstract}
The collapse of marginally bound, inhomogeneous,  pressureless (dust) matter, in spherical symmetry,  is considered. The starting point is not, in this case, the integration of the Einstein equations from some suitable initial conditions. Instead, starting from the corresponding general exact solution of these equations, depending on two arbitrary functions of the radial coordinate, the fulfillment of the Lichnerowicz matching conditions of the interior collapsing metric and the exterior Schwarzschild one is tentatively assumed (the continuity of the metric and its first derivatives on the time-like hypersurface describing the evolution of the spherical 2-surface boundary  of the collapsing cloud), and the consequences of such a tentative assumption are explored. The whole analytical family of resulting models is obtained and some of them are picked out as physical better models on the basis of the finite and constant value of its  {\em intrinsic} energy. 
\end{abstract}

\keywords{Parabolic LTB metrics; marginally bound collapse; intrinsic energy.}

\ccode{PACS numbers: 04.20.Cv, 04.20.-q, 04.20.Jb.}

%\tableofcontents

%%%%%%%%%%%%%
%           %
%  INTRODUCTION %
%           %
%%%%%%%%%%%%%

\section{Introduction: general considerations}
\label{intro}

The subject of this paper is to deal with the implications of reasonable and well motivated simplifying assumptions  in modelling relativistic gravitational collapse scenarios of dust matter in a spherically symmetric  space-time. As it is often done in General Relativity, we assume tentatively that the metric of the differentiable manifold of the theory  is ${\cal C}^1$ class and we will let this assumption to play a central role in the present work by imposing when necessary the corresponding matching Lichnerowicz conditions.\cite{Lichnerowicz} In this context, the novel assumptions we consider and analyze, and whose correctness we verify,  are: 

\begin{enumerate}
\item [(A)] Out of the intrinsic singularity,  there exists a global Gauss coordinate system, adapted to the symmetry and to the dust, in which the metric is of class ${\cal C}^1$, spatially flat and asymptotically Minkowskian, at rest at this 3-space infinity.
\item  [(B)] We require the resulting collapsing models to have finite and constant {\em intrinsic} energy. Then,  additional sound requirements lead finally to a particular simple physical situation. 
\end{enumerate}

More specifically, the interior dust region and the exterior Schwarzschild one should be smoothly matched via the fulfillment of the mentioned Lichnerowicz matching conditions \cite{Lichnerowicz} in the above Gauss coordinates, which are both (i) comoving with the dust at the interior region and (ii) adapted every where to the spherical symmetry and  to free-falling observers at rest at the spatial infinity (parabolic or marginally bound collapse situation). Contrarily, in the above Gauss coordinates satisfying (A) and (B), which will be called {\em intrinsic} coordinates (see Sec. \ref{sec:6}), the well known Oppenheimer-Snyder metric \cite{Oppenheimer-Snyder-1939} does not satisfy neither the Lichnerowicz matching conditions, nor the finite and constant character of the {\em intrinsic} energy. We will see it in Sec. \ref{sec:4} and subsection \ref{sec:6B}. %

Contrarily, some of our collapsing metrics satisfying the above conditions (i) and (ii) have actually vanishing {\em intrinsic} energy, which is a particular way of having a finite constant {\em intrinsic} energy.

But,  we and other authors have elsewhere argued 
\cite{Lapiedra-Morales-2012,Lapiedra-Morales-2013,JuanMari-estable-2014,Ferrando,Lapiedra-Saez} 
that universes having zero {\em intrinsic} energy (and zero linear and angular momenta) could be  good candidates to  be quantically  created from vacuum fluctuations, according to the original Fomin and Tryon suggestions. \cite{Fomin-73,Tryon}

In their pioneering work, Oppenheimer and Snyder \cite{Oppenheimer-Snyder-1939} studied qualitatively  the main issues concerning gravitational collapse scenarios, and they constructed an exact spherically symmetric col·lapse model of a pressureless homogeneous matter with parabolic (also called marginally bound) evolution.  Parabolic evolution means that this Oppenheimer-Snyder model was adapted to a Gauss congruence of observers, comoving with the homogeneous dust fluid, whose simultaneous 3-surfaces of proper constant time are flat (the induced 3-metric has a vanishing Ricci tensor). The interior region is described by the spatially flat Einstein-de Sitter metric and the exterior one by the Schwarzschild solution, both expressed  in Gauss coordinates.%
%%%%FOOTNOTE-1
%
\footnote{ Otherwise, the spherically symmetric homogeneous dust collapse described by a closed Friedmann metric matched with the exterior Schwarzschild solution is extensively analyzed in a lot   of General Relativity text books (see, for instance, Refs. \refcite{Gravitation,Weinberg}). Sometimes, this collapse has been called ``Oppenheimer-Snyder collapse'', though perhaps this name should be better reserved for the ``marginally bound homogeneous'' case considered in  Ref. \refcite{Oppenheimer-Snyder-1939} (and appropriately termed in this manner in Refs. \refcite{Eardley-Smarr-1979,Firouzjaee-Ellis-2015}). Currently,  after the English translation of the work by Datt \cite{Datt-1938} (see also Refs. \refcite{PlebKra,Exact-2003}), the extended appellation ``Oppenheimer-Snyder-Datt collapse''  is also in use, \cite{Joshi-Malafarina-2011,Acquaviva-Ellis-GosHa-2015} although Ref. \refcite{Datt-1938}  exclusively concerns the exact integration of the Einstein equations in some special cases, but not its potential application to gravitational collapse scenarios. \label{Comentari-OP}}
%
%%%%END- FOOTNOTE-1
%
This collapse situation, as well as its inhomogeneous counterpart, although not being quite realistic, are good theoretical models due to their simplicity and provide important information when they are taken as a test-bed in numerical relativity. \cite{Eardley-Smarr-1979}

Therefore, it seems very suitable to extend the Oppenheimer-Snyder spherical marginal bound situation to include an inhomogeneous dust allowing radial non trivial dependence of the proper energy density.  Thus, in the present paper we deal with an inhomogeneous generalization of the result in  Ref. \refcite{Oppenheimer-Snyder-1939} {\em without imposing any particular initial conditions}. However, we tentatively impose in an admissible coordinate system the Lichnerowicz matching conditions. \cite{Lichnerowicz} Thus, the corresponding inner and outer solution are consistently matched on a timelike 3-surface describing the evolution of the star 2-surface collapse. As far as we know, such a study, involving non homogeneous density and matching Lichnerowicz conditions, has never been performed, at least in all the literature quoted in the present paper.

The Lichnerowicz conditions are only well known axioms taken as devices to define global metric solutions. These axioms have the property of being satisfied in the Newtonian approximation, that is, the Poisson equation for the Newtonian potential admits ${\cal C}^1$ solutions for any space bounded, regular, finite mass, source. Of course, even with this validity in the Newtonian approximation, we could deny these axioms in general, but in our manuscript, in the domain of the metric solutions that we have considered, we have found more interesting to assume them tentatively and explore the consequences of the assumption. In any case, see Ref. \refcite{Fayos-Seno-Torres} and references therein for an alternative matching study without requiring necessarily the Lichnerowicz conditions. 

Of course, a lot of work has been carried out in order to obtain exact analytical models for a perfect fluid stellar collapse and to analyze its related issues concerning singularities, event horizons  and trapped surfaces (for a vast bibliography on the subject see, for instance, Refs. \refcite{Joshi-Malafarina-2011,Jos-Dwi,Lasky-2007,Joshi-Malafarina-Narayan-2014,LaMo-2017}). This extensive work may be considered as an extension of the Oppenheimer-Snyder collapse by including density inhomogeneities and pressure contributions, and has been mainly focussed  on the role played by the initial conditions and the subsequent collapsing evolution. Nevertheless, here, keeping on the specific case of a marginally bound spherically symmetric dust collapse, we will exhaustively analyze, as remarked above,  the role played by the boundary conditions imposed on the interior-exterior matching surface without imposing initial conditions.

The paper obeys the following structure. First, in Sec. \ref{sec:2} we start from the known expression for the LTB (Lema{\^{\i}}tre,\cite{Lemaitre-1933} Tolman,\cite{Tolman-1934} Bondi\cite{Bondi-1947}) family of exact solutions in Gauss comoving coordinates adapted to the spherical symmetry, in the marginal bound case (i. e. when the three-space curvature tensor vanishes), from which one recovers the Schwarzschild metric as a particular case. In Sec.  \ref{sec:3} we construct a family of solutions of the Einstein equations that is of ${\cal C}^1$ class over its whole domain and that describes a collapsing inhomogeneous dust fluid matched to the Schwarzschild exterior written in the Gauss coordinate system introduced in Sec. \ref{sec:2}, and obeying the marginally bound condition. The homogeneous Oppenheimer-Snyder dust collapse \cite{Oppenheimer-Snyder-1939} situation is revisited in Sec. \ref{sec:4}, and then, this important model is contrasted with the generic matching model presented in Sec. \ref{sec:3}. Then, for the sake of clarity, in Sec. \ref{sec:5} a {\em truncated} one-parametric matching metric family is introduced in order to examine the minimal set of additional physical requirements allowing to fix this parameter. In Sec. \ref{sec:6} we show that this parameter can  be naturally determined by requiring that the {\em intrinsic} energy of the whole matching solution be finite and constant. In the ending Sec. \ref{sec:7} we briefly discuss and comment the obtained results. An appendix presents the mathematics concerning the application of the Lichnerowicz matching conditions \cite{Lichnerowicz} to the generic dust collapse analyzed in Sec. \ref{sec:3}.

We will take units in which $c=1$ for the speed of light, and $G=1$ for the gravitational constant. The metric signature will be $+2$.

A short communication containing some results, without proof,  of this
work has been exposed at the Spanish Relativity Meeting ERE-2015. %\cite{ere2015}.

%
%
%

%%%%%%%%%%%%%
%           %
%  SECTION  2  %
%           %
%%%%%%%%%%%%%

\section{Schwarzschild metric in suitable Gauss coordinates}
\label{sec:2}

Let it be the  Schwarzschild metric in standard static coordinates:
\begin{equation}\label{Smetrica}
 g_{_S}= - \Big(1 - \frac{2m}{r}\Big)dt \otimes dt  +  \frac{dr \otimes dr}{\displaystyle{1 - \frac{2m}{r}}} + r^2 h, 
\end{equation}
with  $h = d\theta \otimes d\theta + \sin^2 {\hspace{-0.7mm}} \theta \, d \phi \otimes d\phi$ the metric on the unit 2-sphere, and $m$ the source mass parameter.
Let us change to the new coordinates $(\tau, \rho)$ given implicitly by
\begin{equation} \label{canvi}
\tau (t, r) = t + 2 m \Big[2 \sqrt{\frac{r}{2m}} + \ln \Big|\frac{\sqrt{r} - \sqrt{2m}}{\sqrt{r} + \sqrt{2m}}\Big|\Big], 
\end{equation}
\begin{equation} \label{rhoR}
r (\tau, \rho) = \Big[\frac{9}{2} m \Big(\tau - \frac{2}{3} \frac{\rho^{3/2}}{\sqrt{2m}}\Big)^2\Big] ^{1/3}, \quad \tau < \frac{2}{3} \frac{\rho^{3/2}}{\sqrt{2m}}.
\end{equation}

In the new coordinates still adapted to the spherical symmetry of the metric (\ref{Smetrica}) we have:
\begin{equation}\label{LemaitreP}
g_{_S}  =  - d\tau \otimes d\tau +  \frac{\rho}{r} d\rho \otimes d\rho+ r^2 h, 
\end{equation}
which is a regular extension of the Schwarzschild metric (\ref{Smetrica}) to the region $r\in(0, +\infty)$, describing  the black hole region with its horizon, $r \in(0, 2m]$, and its exterior $r > 2m$. In Kruskal coordinates \cite{Kruskal-1960} $\{u, v, \theta, \phi\}$, this whole region is $u>v$ and arbitrary $\theta$ and $\phi$.

The above metric form (\ref{LemaitreP}) of the Schwarzschild solution has been considered in Refs. \refcite{Oppenheimer-Snyder-1939,Lapiedra-Morales-2013,Synge-1950,Robertson-Noonan}. It is written in Gauss coordinates, that is,  the metric components $g_{00}$ and  $g_{0i}$ $(i = 1, 2, 3)$ are $g_{00} = -1$, $g_{0i} = 0$, and they become asymptotically at rest coordinates for $\rho \to \infty$, with respect to ordinary static Schwarzschild coordinates, since on the one hand we have $\rho/r \sim 1 + \sqrt{2m} \, \tau \rho^{-3/2} \to 1$, and $\tau$ is an up bounded variable: notice that, according to Eq. (3),  where $\psi(\lambda) \equiv 2 \lambda^{3/2}/(3 \sqrt{2m})$ is the singular time corresponding to the collapse of the star surface $\rho = \lambda$. Besides, on the other hand, we have from (\ref{canvi}), for $r \to \infty$, $d \tau  \sim dt + \sqrt{2m/r} \sim dt$. Furthermore,  the coordinates $\{\tau, \rho\}$ are regular coordinates everywhere, except for the essential singularity $r = 0$ and, with the appropriate coordinate transformation,  they become asymptotically rectilinear at the three-space infinity 
(for $\rho = r$, metric (\ref{LemaitreP}) becomes the Minkowski metric in spherical coordinates). The metric form (\ref{LemaitreP}) is a very suitable one for ascribing an {\em intrinsic} energy to the space-time metric describing a marginally bound gravitational collapse (see Sec. \ref{sec:6}).%
%%%%FOOTNOTE-2
%
\footnote{Another appropriated and well known Gauss coordinate system for the Schwarzschild space-time is the one introduced by Novikov \cite{Novikov} (see also Ref. \refcite{Gravitation}).  In this case, the three-spaces $\tau = constant$ are not flat, and then, in dealing with a marginally bound collapse scenario,  the associated Novikov metric form is not as suitable as the metric form (\ref{LemaitreP}).  \label{Comentari-Novikov}}
%
%%%%END- FOOTNOTE-2
%
 This Schwarzschild metric is a particular case of the LTB metric, solution of the Einstein equations (see, for instance, Refs. \refcite{PlebKra,Exact-2003}): 
\begin{equation}\label{LTB0}
g =-d\tau \otimes d\tau + A'^{\,2} d\rho \otimes d\rho +A^2 h ,  
\end{equation}
where $A' \equiv \partial_\rho A$, and $A \equiv A (\tau,\rho)$ is given by
\begin{equation} \label{k0}
A(\tau, \rho) = \displaystyle{\Big\{\frac{9}{2} M(\rho) [\tau - \psi(\rho)]^2\Big\}^{1/3}}, 
\end{equation}
$M(\rho)$ and $\psi(\rho)$ being arbitrary functions of $\rho$, $0 \leq \rho < \infty$. The time coordinate domain is $- \infty  < \tau  < \psi (\rho)$. In particular, taking
\begin{equation} \label{Mrho}
M(\rho) = const. = m > 0,  \quad  \psi(\rho) =  \frac{2}{3} \frac{\rho^{3/2}}{\sqrt{2m}}, 
\end{equation}
we recover (\ref{LemaitreP}), with $A (\tau, \rho)= r$ given by Eq. (\ref{rhoR}). Let us consider in this particular case the times $\tau_h$ and $\tau_s$ defined respectively by
$A(\tau_h, \rho) = 2m$ and $A(\tau_s, \rho) = 0$ (with $\tau_s > \tau_h$), that is
\begin{equation} \label{taus-h}
\tau_h = \tau_s -\frac{4m}{3},  \qquad \tau_s = \frac{2}{3} \frac{\rho^{3/2}}{\sqrt{2m}}.
\end{equation}
A free falling observer whose world line is given by constant $\rho, \theta$ and $\phi$ reaches the event horizon $r=2m$ at proper time $\tau_h$, and the essential singularity 
$r=0$ at $\tau_s > \tau_h$.

The metric (\ref{LTB0}) involving two arbitrary functions  of $\rho$,  $M(\rho)$ and $\psi(\rho)$, is sourced by a pressureless (dust) matter content  whose stress energy tensor $T^{\alpha\beta}$ is
\begin{equation}\label{dust}
T^{\alpha\beta}= \mu(\tau, \rho) u^\alpha u^\beta, 
\end{equation}
with $\mu(\tau, \rho)$ the proper dust density function, which is given by:
\begin{equation}\label{mu-dust}
4 \pi \mu = \frac{M'}{A^2 A'} 
\end{equation}
(see again Refs. \refcite{PlebKra,Exact-2003} for a brief account concerning the related whole family of exact solutions). Notice that the above $(\tau, \rho, \theta, \phi)$ Gauss coordinates are comoving  with the source dust and are adapted to the present spherical symmetry, that is,  in particular,  $u^\alpha = (1, 0, 0, 0)$.

Function $M(\rho)$ can be seen to be the total mass inside the radius $\rho$, and is the particular value that, in our case, the Misner-Sharp \cite{Misner-Sharp} mass function takes. Function $\psi(\rho)$ represents the proper time when 
the essential singularity $A(\tau, \rho) = 0$ takes place. More precisely, let it be a radially falling observer at constant radial comoving coordinate $\rho$ from  the center $\rho =0$; then its world line meets the essential singularity at the proper time $\tau_s = \psi(\rho)$. 

Of course, a gauge freedom remains: the radial coordinate changes $\bar{\rho} = \bar{\rho} (\rho)$ leaving invariant the metric form (\ref{LTB0}). Then,  one of the free functions $M(\rho)$ or $\psi(\rho)$  may be equalled  to an arbitrarily prescribed function of $\rho$. Actually, as we will see below in detail, for a collapse model, the required matching conditions constrain the values that functions $M(\rho)$ and $\psi(\rho)$, and their derivatives, can take on the matching surface. This is a crucial consideration for the interpretation of our results: in our case, the set of conditions that $M(\rho)$ and $\psi(\rho)$ must fulfill are required by the accomplishment of the {\em Lichnerowicz matching}  conditions  \cite{Lichnerowicz} (the matching of the interior space-time metric, describing the collapse, and the  Schwarzschild exterior). We will come back to this point below (see Sec. \ref{sec:4}) in connection with the familiar prescription of the function $\psi(\rho)$ usually taken for marginally bound homogeneous dust collapse (see, for instance, Refs. \refcite{Oppenheimer-Snyder-1939,Eardley-Smarr-1979}).

As mentioned above, metric (\ref{LTB0}) is a subclass of the LTB family of dust solutions, where the abbreviation comes from the original works by Lema{\^{\i}}tre \cite{Lemaitre-1933}, Tolman \cite{Tolman-1934} and Bondi \cite{Bondi-1947}. This subclass is the most general spherical symmetric dust solution with flat three-spaces given by $\tau =constant$.%
%%%%FOOTNOTE-3
%
\footnote{In fact, in the case of the  Datt metric $g_{_D}$ (see Ref. \refcite{Datt-1938,PlebKra,Exact-2003}) the three-spaces $\tau = constant$ are non-flat. Moreover, these 3-spaces are the level surfaces of the $A$ function of this solution because it is a function of the sole proper time $\tau$, $A(\tau)$. This means that the gradient $dA$ is everywhere timelike, $g_{_D}(dA,dA) <0$,  on the whole metric domain (which is a $T$-region, \cite{Novikov} i. e, the Datt function $A$ defines a timelike gradient coordinate according to the parlance used in Refs. \refcite{4Nw-CoFeMo-2009,199Lz-CoFeMo-2009}).  \label{Comentari-Datt}}
%
%%%%END- FOOTNOTE-3
%

%%%%%%%%%%%%%
%           %
%  SECTION  3 %
%           %
%%%%%%%%%%%%%

\section{Generic parabolic matching model for inhomogeneous dust collapse}
\label{sec:3}

Assume, as it is often done, that the differentiable manifold, $V_4$, of General Relativity, is of ${\cal C}^2$ class and that the gravitational potentials are of ${\cal C}^1$ class, i.e., the metric first derivatives are locally continuous functions, that is to say, are continuous everywhere the coordinate system used exists. Can we match the Schwarzschild exterior metric form (\ref{LemaitreP}) with an interior one represented by (\ref{LTB0}) with $A(\tau, \rho)$ given by  (\ref{k0})? We will see that such is the case.%
%
%%%%FOOTNOTE-4
%
\footnote{Notice that, according to Ref. \refcite{Lichnerowicz}, these differentiability requirements on the  gravitational potentials are manifestly compatible with the differential structure of $V_4$ if, in addition, one requires that the Hessian matrix components of the admissible coordinate transformations  are piece-wise ${\cal C}^2$ class functions (or, ``par morceaux'' according to the nomenclature used in  Ref. \refcite{Lichnerowicz}). This condition ensures that,  under  coordinates changes whose third and  fourth derivatives could present  some Hadamard discontinuities (see Ref. \refcite{Hadamard} p. 85, and Eq. (\ref{H-dis}) next), the ${\cal C}^1$ differential character of the metric components remains invariant. Nevertheless, the second derivatives of the metric could present some Hadamard discontinuities (the metric first derivatives being at last piece-wise ${\cal C}^2$ class). See Ref. \refcite{Lake-2017} for a recent discussion concerning these requirements. \label{sobre-Licnne}}
%
%%%%END- FOOTNOTE-4
%

To begin with, since $\rho$ in  (\ref{LemaitreP}) is a comoving radial  3-space coordinate, everywhere out of $r=0$, let it be $\rho = const. \equiv \lambda$, the value of this coordinate for the corresponding 2-surface separating the outside region from the inside one, at the time $\tau$. Then, in accordance with the general Lichnerowicz prescription, \cite{Lichnerowicz} expressing in this case the assumed ${\cal C}^1$ character of the metric, the 3-space metric components $g_{ij}$ ($g_{00}=-1$, $g_{0i}=0$, everywhere out of $r=0$), and their first derivatives, have to be continuous across $\rho = \lambda$, in order to consistently match both solutions, the exterior Schwarzschild metric with a particular  interior LTB solution (Eq. (\ref{LTB0})) driven by a pressureless matter, i.e., by dust.

By derivation of (\ref{k0}) with respect to $\tau$ and $\rho$,  the following expressions result:
\begin{equation}\label{A-punt}
\dot{A} = \varepsilon \sqrt{\frac{2M}{A}}, 
\end{equation}
\begin{equation}\label{A-prima}
A' = \frac{1}{3} \frac{M'}{M} \, A - \varepsilon \sqrt{\frac{2M}{A}} \, \psi', 
\end{equation}
with $\varepsilon \equiv sgn[\tau - \psi(\rho)]$ describing, respectively, an expanding ($\dot{A} > 0$) or contracting  ($\dot{A} < 0$) configuration when $\varepsilon = +1$ or $\varepsilon =-1$. Then, taking into account that $A^{\frac{3}{2}} (\tau, \rho)= \varepsilon \,  \frac{3}{2}\, \sqrt{2M(\rho)} \,  [\tau-\psi(\rho)]$, the events $(\tau, \rho, \theta, \phi)$ where $A'$ vanishes display generically a 3-surface given by  the relation:
\begin{equation}\label{A-prima-zero}
A'(\tau, \rho) = 0 \iff \tau (\rho) = \psi(\rho) + 2 \frac{M(\rho)}{M'(\rho)} \psi'(\rho)
\end{equation}
and arbitrary $(\theta, \phi)$. This relation can be interpreted as follows.  Assuming that $M(\rho)$ is a non-negative function, the proper time $\tau(\rho)$ at which $A'$ vanishes will be hidden under the essential singularity at $A=0$ (that is $\tau(\rho) > \psi(\rho)$) if, and only if, $M'(\rho)$ and  $\psi'(\rho)$ have equal sign, i.e. if they are both increasing or decreasing functions. The question of this hiddeness possibility is the problem named of shell crossing. See Refs. \refcite{Hellaby-Lake-85,Hellaby-Lake-86}.

Then, for a gravitational collapse ($\varepsilon =-1$)  the reader can easily verify the above matching can be achieved with the following values of the functions $M(\rho)$ and $\psi(\rho)$:
\begin{equation} \label{Mrho2} 
M(\rho) = \, \left \{
\begin{array} {ll}
\displaystyle{m +  \sum_{k=3}^\infty M_k (1- \frac{\rho}{\lambda})^k, \quad \rho \leq  \lambda} \\
m, \quad \rho \geq \lambda 
\end{array} \right.
\end{equation}
and 
\begin{equation} \label{psi} 
\psi(\rho) = \, \left \{
\begin{array} {ll}
\displaystyle{\frac{\lambda^{3/2}}{4 \sqrt{2m}} (\frac{\rho^2}{\lambda^2} + 2 \frac{\rho}{\lambda}-\frac{1}{3})+  \sum_{k=3}^\infty \psi_k (1- \frac{\rho}{\lambda})^k,  \rho \leq  \lambda} \\
\displaystyle{\frac{2}{3} \frac{\rho^{3/2}} {\sqrt{2m}}, \quad \rho \geq \lambda}
\end{array} \right.
\end{equation}
giving what we call the ``generic matching metric'' (see \ref{ap-A}),  where the constant coefficients $M_k$ and $\psi_k$, otherwise arbitrary,  have to ensure the convergence of the corresponding series, and will be constrained  by the remaining physical requirements imposed on  the model 
(see Secs. \ref{sec:5} and \ref{sec:6}).

Functions  (\ref{Mrho2})  and  (\ref{psi}) form a broad class of analytic solutions for our matching problem.%
%%%FOOTNOTE-5
%
\footnote{Note that the required matching conditions are also satisfied by taking for $\psi(\rho)$ the generalization of the  expression (\ref{Mrho}) we have chosen for the Schwarzschild metric, that is, $\psi(\rho) \to \widetilde\psi(\rho)\equiv \frac{2}{3} \frac{\rho^{3/2}} {\sqrt{2M(\rho)}}$ for all $\rho >0$, with $M(\rho)$ given by (\ref{Mrho2}). This means that for $\rho \leq \lambda$ the general family of $\psi(\rho)$ functions obeying the matching requirement is a weighted linear combination $a\psi(\rho)+ (1-a) \widetilde\psi(\rho)$ with  $\psi(\rho)$ given by (\ref{psi}), and $a\in[0,1]$. A model with  $a=0$ has been recently analyzed and applied to test the fulfillment of the cosmic censorship conjecture in a broad class of inhomogeneous dust collapsing scenarios (see Ref. \refcite{LaMo-2017}). \label{sobre-psi}}
%
%%%%END- FOOTNOTE-5
%
This solution, for $\rho \geq \lambda$, reproduces the exterior Schwarzschild metric in the coordinates of (\ref{LemaitreP}) with $r \equiv r(\tau, \rho)$ given by (\ref{rhoR}). Obviously, any of these solutions could also be obtained by integrating the Einstein field equations with the corresponding initial conditions $A(\tau_i, \rho)$,  $\dot A(\tau_i, \rho)$, $\tau_i$ standing for the initial time $\tau_i - \psi (0)<0$. The particular values of the expressions 
$A(\tau_i, \rho)$,  $\dot A(\tau_i, \rho)$, would be obtained substituting in (\ref{k0}) and (\ref{A-punt}), respectively, the corresponding values of $M$ and $\psi$ given by (\ref{Mrho2}), (\ref{psi}), which insures the ${\cal C}^1$ class character of these initial conditions. 

It is important to stress that, in accordance with the general result in Ref. \refcite{Bonnor-Vickers},  the Darmois--Israel matching conditions \cite{Darmois1927,Israel-1966} (continuity of the induced metric and the extrinsic curvature across the matching $\Sigma$ surface $\rho = \lambda$)  are also fulfilled by the space-time metric (\ref{LTB0}) no matter how the  $A(\tau, \rho)$ function given by (\ref{k0}) may be, provided that $A$ be continuous across the matching surface $\Sigma$. In fact, the induced metric, $\gamma$, and  the extrinsic curvature, $K$,  of a time-like three-surface defined by $\rho =constant$ are given by  
\begin{equation}\label{gamma-K}
\gamma = - d\tau \otimes d\tau + A^2(\tau, \rho) h, \quad K = - A(\tau, \rho) h.
\end{equation}
Then, the continuity of $A$ across the matching surface $\Sigma$,  expressed like 
$[A]_{_\Sigma} = 0$, ensures the continuity of both $\gamma$ and $K$ across $\Sigma$, that is, in the same notation,  $[\gamma]_{_\Sigma} = [K]_{_\Sigma} = 0$ are indeed fulfilled.

For $\tau = 0$ and for all $\rho \geq \lambda$, the substitution of (\ref{Mrho2}) and (\ref{psi}) into (\ref{k0}) and (\ref{A-punt}) leads to $A(0, \rho) = \rho$ and $\dot{A}(0, \rho) = - \sqrt{\frac{2m}{\rho}}$. These are then the initial conditions corresponding to our ``generic matching metric''  obtained from (\ref{Mrho2}) and (\ref{psi}), for $\rho \geq \lambda$. In fact, this last initial condition says that, at each $\rho \geq \lambda$, the speed of the infalling observer is equal to the Newtonian escape velocity. But,  such initial values are not always satisfied for $\rho <\lambda$ in our model (\ref{Mrho2}) and (\ref{psi}) due to the fact that this model has been constructed by requiring the accomplishment of the Lichnerowicz matching conditions.

This is a characteristic of the present models, making them different from other models considered in the literature which imposes that, under a global radial coordinate choice,  the above initial conditions, $A(0, \rho) = \rho$ and $\dot{A}(0, \rho) = - \sqrt{\frac{2M(\rho)}{\rho}}$ be satisfied for every value of $\rho \in(0, \infty)$ (cf. Refs. \refcite{Joshi-Malafarina-2011,Lasky-2007}).

%%%%%%%%%%%%%
%           %
% SECTION 4 %
%           %
%%%%%%%%%%%%%

\section{Oppenheimer-Snyder homogeneous dust collapse}
\label{sec:4}

In Ref. \refcite{Oppenheimer-Snyder-1939}, the marginally dust homogeneous spherically symmetric collapse was considered by expressing the 
functions $M(\rho)$ and  $\psi(\rho)$ (after a change in notation) in the following manner:

\begin{equation} \label{Mrho-OS} 
M_{_{\rm{OS}}}(\rho) = \, \left \{
\begin{array} {ll}
\displaystyle{m \Big(\frac{\rho}{\lambda}\Big)^3, \quad \rho \leq  \lambda} \\
m, \quad \rho \geq \lambda 
\end{array} \right.
\end{equation}
and 
\begin{equation} \label{psi-OS} 
\psi_{_{\rm{OS}}}(\rho) = \, \left \{
\begin{array} {ll}
\displaystyle{\frac{2}{3} \frac{\lambda^{3/2}}{\sqrt{2m}},   \quad  \rho \leq  \lambda}\\ \\
\displaystyle{\frac{2}{3} \frac{\rho^{3/2}}{\sqrt{2m}}, \quad \rho \geq \lambda}, 
\end{array} \right.
\end{equation}
where we use the OS abbreviation for Oppenheimer and Snyder. \cite{Oppenheimer-Snyder-1939} In this case, the areal radial metric function $A(\tau, \rho)$ is
\begin{equation} \label{A-function-OS} 
A_{_{\rm{OS}}}(\tau, \rho) = \, \left \{
\begin{array} {ll}
\displaystyle{\Big(\frac{9m}{2}\Big)^{1/3} \frac{\rho}{\lambda} \Big(\tau - \frac{2}{3} \frac{\lambda^{3/2}}{\sqrt{2m}}\Big)^{2/3},  \,  \,  \rho \leq  \lambda}\\ \\
\displaystyle{\Big(\frac{9m}{2}\Big)^{1/3} \Big(\tau - \frac{2}{3} \frac{\rho^{3/2}}{\sqrt{2m}}\Big)^{2/3},  \quad  \rho \geq  \lambda}. 
\end{array} \right.
\end{equation}
As it has been stressed at Sec. \ref{sec:3},  about Eq. (\ref{gamma-K}), the Darmois matching conditions are also fulfilled in the present case of an homogeneous (see next Eq. (\ref{mu-dust-homo})) collapse, whose space-time metric is (\ref{LTB0}) with  $A(\tau, \rho)$ given by  (\ref{A-function-OS}). This is so since 
$[A_{_{\rm{OS}}}]_{_\Sigma} = 0$. However $[A'_{_{\rm{OS}}}]_{_\Sigma} \neq 0$,  because
\begin{equation}\label{A-prima-OS-dis}
[A'_{_{\rm{OS}}}]_{_\Sigma} = - \frac{q}{\lambda}\, \frac{\tau}{(\tau - \tau_c)^{1/3}},%
\end{equation}
where for a given function $X(\tau,\rho)$, its Hadamard discontinuity \cite{Hadamard}, $[X]_{_\Sigma}$,  across the matching $\Sigma$ surface $\rho = \lambda$,  is defined by 
\begin{equation}\label{H-dis}
[X]_{_\Sigma} = \lim_{\epsilon \to 0^+} X(\tau, \lambda+\epsilon) - \lim_{\epsilon \to 0^+} X(\tau, \lambda - \epsilon), 
\end{equation}
and, where the notation 
\begin{equation}\label{qtauc}
q \equiv \Big(\frac{9m}{2}\Big)^{1/3},  \quad  \tau_c \equiv \frac{2}{3} \frac{\lambda^{3/2}}{\sqrt{2m}},
\end{equation}
will be used from now on, $\tau_c$ being the time when the ``simultaneous'' essential singularity $A_{_{\rm{OS}}}(\tau_s, \rho)= 0$ 
occurs for every dust comoving observer (again, see next Eq. (\ref{mu-dust-homo})).

\begin{figure}
\centerline{
\parbox[c]{0.7\textwidth}{\includegraphics[width=0.6\textwidth]{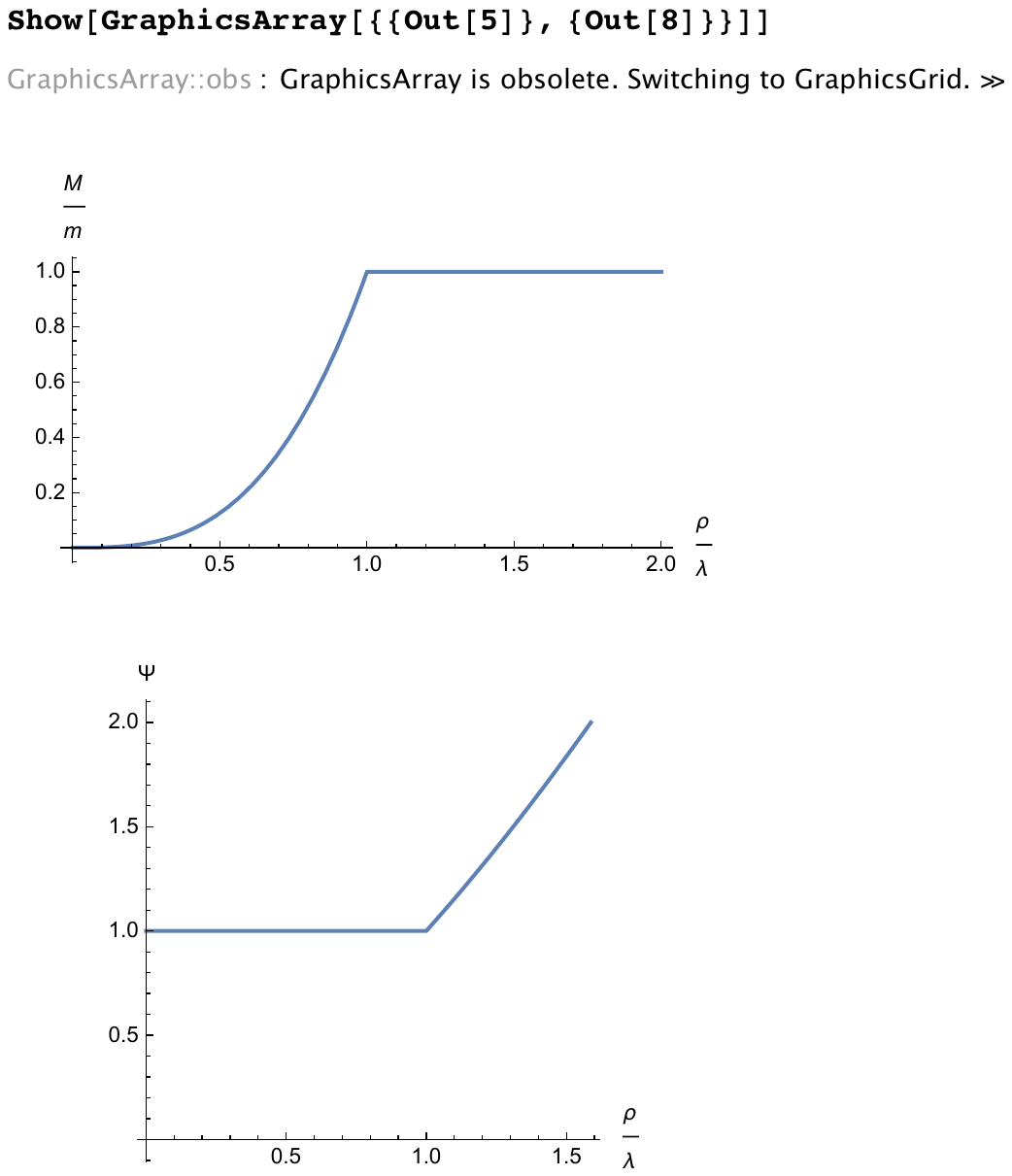}}} 
\caption{$M_{_{\rm{OS}}}(\rho)$ (top) and $\psi_{_{\rm{OS}}}(\rho)$ (bottom) functions describing the Oppenheimer-Snyder collapse. 
Their first derivative are discontinuous across the matching surface  $\{\Sigma: \rho = \lambda$\}, implying that $A'_{_{\rm{OS}}}$ is also discontinuous across $\Sigma$, 
$[A'_{_{\rm{OS}}}]_{_\Sigma}\neq 0$. Normalized variable, $\frac{\rho}{\lambda}$, and  normalized functions,  
$\frac{M}{m}\equiv \frac{M_{_{\rm{OS}}}}{m}$ and $\Psi \equiv \frac{3}{2}\frac{\sqrt{2m}}{\lambda^{3/2}} \psi_{_{\rm{OS}}}$, are used. \label{OS-collapse}}
\end{figure}

Then,  the metric component $g_{\rho\rho} = A'^{\, 2}$ is discontinuous across $\Sigma$, 
\begin{equation}\label{A-prima-quadrat-OS-dis}
[A'^2_{_{\rm{OS}}}]_{_\Sigma} = - \frac{q^2}{\lambda^2} \,  \frac{\tau (\tau - 2 \tau_c)}{ (\tau - \tau_c)^{2/3}}.
\end{equation}

Consequently, the tentatively assumed differentiability  axioms \cite{Lichnerowicz} requiring that the gravitational potentials $g_{\mu \nu} (x^\alpha)$ be functions of class ${\cal C}^1$ (i. e. continuously differentiable functions of the admissible coordinates $\{x^\alpha\}$) are not fulfilled in this case (see Figure \ref{OS-collapse}).

Of course, since the Oppenheimer-Snyder metric fulfills the matching Darmois conditions through $\Sigma$, there are local Gauss coordinate systems built from $\Sigma$ where the matching Lichnerowicz conditions of this metric are fulfilled too. This is actually the specific meaning, referred to the present case, of the general result proved in Ref. \refcite{Bonnor-Vickers}, that under some differentiable requirements, fulfilled in our case, Darmois and Lichnerowicz conditions are ``equivalent''. However, these local coordinates will not be anymore {\em intrinsic} coordinates as they are in the case of the general metrics of Sec. \ref{sec:3}. Actually, the sole freedom we are left in the coordinate choice is the one related to the gauge freedom $\rho \to \bar{\rho}(\rho)$, mentioned in the penultimate paragraph of Sec. \ref{sec:2}. Obviously, this kind of freedom cannot account for the time depending metric discontinuity present in Eq. (\ref{A-prima-quadrat-OS-dis}). These {\em intrinsic} coordinates, whose uniqueness will be discussed next, at the end of the subsection  \ref{sec:6A}, are the appropriate ones to calculate the {\em intrinsic} energy of the metric considered (see Sec. \ref{sec:6}). We want to calculate this {\em intrinsic} energy, for the metrics of the present paper, in order to discriminate those of these metrics that would seem specially physical (the ones with a constant and finite {\em intrinsic} energy), from the remaining ones (with a time dependent {\em intrinsic} energy, that even more could be infinite).

Substituting  (\ref{Mrho-OS}) and (\ref{A-function-OS}) for $\rho < \lambda$ in  (\ref{mu-dust}), the dust density becomes homogeneous  in this case:
\begin{equation}\label{mu-dust-homo}
\mu_{_{\rm{OS}}} (\tau) = \displaystyle{\frac{1}{6 \pi  (\tau - \tau_c)^2}},   
\end{equation}
and approaches to infinite  when  $\tau$ goes to  $\tau_c$.  The 3-surface  (apparent horizon) defined by the equation $g(dA_{_{\rm{OS}}}, dA_{_{\rm{OS}}}) =0$, that is equivalent to the algebraic condition $A_{_{\rm{OS}}}(\tau, \rho)= 2M_{_{\rm{OS}}}(\rho)$,  is reached by the comoving observers at the time $\tau_h (\rho)< \tau_s$
given by
\begin{equation}\label{tau-h-OS}
\tau_h(\rho) = \tau_c - \frac{4 m}{3} \Big(\frac{\rho}{\lambda}\Big)^3
\end{equation}
for $\rho \leq \lambda$, while for $\rho \geq \lambda$ we must write (\ref{taus-h}) instead of (\ref{mu-dust-homo}). Notice that, according to 
(\ref{mu-dust-homo}), the ``central observer'' world-line ($\rho = 0$ and variable $\tau$) belongs to a regular state, except 
for $\tau = \tau_c$ where $\mu_{OS}$ becomes infinite.

%

%%%%%%%%%%%%%
%           %
% SECTION  5 %
%           %
%%%%%%%%%%%%%

\section{Truncated matching metric for inhomogeneous dust collapse}
\label{sec:5}

In this section we construct a {\em truncated}  solution for a dust collapse satisfying the differentiability and matching requirements, 
according with the prescriptions considered in the Sec. \ref{sec:3}, and then, we contrast its main characteristics with the more familiar 
ones founded in the literature on the subject. Such a model is defined by the following specification of the free functions $M(\rho)$ and $\psi(\rho)$, 
\begin{equation} \label{Mrho-simple} 
M(\rho) = \, \left \{
\begin{array} {ll}
\displaystyle{m -  \alpha \Big(1- \frac{\rho^2}{\lambda^2}\Big)^3, \quad \rho \leq  \lambda} \\ 
m, \quad \rho \geq \lambda 
\end{array} \right.
\end{equation}
and 
\begin{equation} \label{psi-simple} 
\psi(\rho) = \, \left \{
\begin{array} {ll}
\displaystyle{\frac{\lambda^{3/2}}{4 \sqrt{2m}} (\frac{\rho^2}{\lambda^2} + 2 \frac{\rho}{\lambda}-\frac{1}{3}), \quad  \rho \leq  \lambda}\\ \\
\displaystyle{\frac{2}{3} \frac{\rho^{3/2}}{\sqrt{2m}}, \quad \rho \geq \lambda},
\end{array} \right.
\end{equation}
and where $\alpha$ is a constant to be determined from additional physical requirements on the model (see next section). Notice that Eq. (\ref{Mrho-simple}) is a special polynomial case that follows from the general development Eq. (\ref{Mrho2}) by considering a particular election of 
the coefficients $M_k$. Moreover,  the fact that Eq. (\ref{Mrho-simple}) only contains even potential exponents, ensures that $M(\rho)$ has also zero first derivative at the center $\rho=0$, according with the regularity conditions imposed in collapse scenarios (see, for instance,  Refs.\refcite{Joshi-Malafarina-2011,Lasky-2007}, and Figure \ref{simple-collapse} in Sec. \ref{sec:6}).

These functions, $M(\rho)$ and $\psi(\rho)$ satisfy
\begin{equation}\label{LichneR}
[M]_{_\Sigma} = [M']_{_\Sigma} = [M'']_{_\Sigma} = [\psi]_{_\Sigma} = [\psi']_{_\Sigma} =[\psi'']_{_\Sigma} = 0, 
\end{equation}
ensuring the fulfillment of the Lichnerowicz matching conditions (\ref{LichneR}), according with the results displayed in the \ref{ap-A}.

For this truncated model, the corresponding ``truncated matching metric''  has an essential singularity $A(\tau, \rho)=0$ which is ``non-simultaneous'' for the dust comoving observers since it occurs at  different proper time values, $\tau_s(\rho)= \psi(\rho)$,  for different observers, that is, 
\begin{equation}\label{sin-ese} 
\tau_s(\rho) = \displaystyle{\frac{\lambda^{3/2}}{4 \sqrt{2m}} \Big(\frac{\rho^2}{\lambda^2} + 2 \frac{\rho}{\lambda}-\frac{1}{3}\Big)}, 
\end{equation}
the density $\mu$ given by (\ref{mu-dust}) becoming infinite at this time.

A natural physical requirement is to take $M(\rho) \geq 0$ and $M'(\rho) > 0$ for $\rho \neq 0$. This is accomplished by taking $\xi \in(0, 1]$, where $\xi\equiv \alpha/m$. Then, taking into account that in (\ref{psi-simple})  $\psi(\rho)$ is an increasing function, $\psi'(\rho)>0$, one gets  from (\ref {A-prima-zero}) that $A'(\tau, \rho)$ can only vanish for $\tau(\rho) \geq \psi(\rho)$. Then, the points $(\tau, \rho)$ where $A'$ vanishes, if any, are hidden by the essential singularity $A=0$.

The function $A(\tau, \rho)$ giving the metric (\ref{LTB0}) for the  model is directly obtained from (\ref{Mrho-simple}) and (\ref{psi-simple}). 
Then, for $\rho \geq \lambda$,  the function $A(\tau, \rho)$ reduces to
$A(\tau, \rho) = r(\tau, \rho)$ indeed, the space-time metric being the  Schwarzschild one in this exterior region, 
and, in particular,  $A(0, \rho \geq \lambda) = \rho$, according to Eq. (\ref{rhoR}). 

An appropriated election we could take for the arbitrary constant $\alpha$ seems to be $\alpha= m$, giving the maximum value for the $\xi$ parameter,  $\xi = 1$. In the next section we justify  such an election from the study of the ADM (Arnowit-Deser-Misner \cite{ADM-1962}) energy associated with the present model, when using {\em intrinsic} coordinates.

%
%

%%%%%%%%%%%%%
%           %
%  SECTION  6 %
%           %
%%%%%%%%%%%%%

\section{Constraining the matched metric from the study of its {\em intrinsic} energy}
\label{sec:6}

In Ref. \refcite{LaMo-2014}, using {\em intrinsic} coordinates, we have studied the metric {\em intrinsic} energy derived from the Weinberg complex prescription \cite{Weinberg} for the general family of LTB-metrics, given by the line element in suitable Gauss coordinates
\begin{equation} \label{metric-LTB}
g_{_{LTB}}=-d\tau \otimes d\tau +\frac{A'^{\, 2}}{1-k(\rho)}d\rho \otimes d\rho +A^2 h, \quad A \equiv A(\tau, \rho)
\end{equation}
in the frame of some particular void models constructed with this family. The function $k(\rho)$ appearing in (\ref{metric-LTB}) is related to the curvature of the proper time synchronization attached to the Gauss observers comoving with the dust fluid. In this case,  and using comoving Gauss coordinates, the calculation of the $Q^{i00}$ component of the Weinberg pseudotensor,  $Q^{i00} = (\partial_i g - \partial_j g_{ij})/2$,  with $g_{ij}$ the three-space metric and $g \equiv \delta^{ij} g_{ij}$,  gives (in accordance with Ref. \refcite{Virbhadra})
\begin{equation}\label{Qi00}
Q^{i00}=-\frac{1}{\rho^2}\Big[\frac{(A- \rho A')^2}{\rho}+\frac{k \rho A'^{\, 2}}{1-k}\Big] n^{i}, 
\end{equation}
where $k \equiv k(\rho)$ and $n^{i} = n_i \equiv \rho_{i}/\rho$ stands for the unit vector associated with  the asymptotic rectilinear coordinates $\rho_i$,  
\begin{equation}\label{rho-cartesianes}
\rho_1 = \rho \sin \theta \cos \phi, \, \rho_2 = \rho \sin \theta \sin \phi, \,  \rho_3 = \rho \cos \theta, 
\end{equation}
constructed from the starting $(\tau, \rho, \theta, \phi)$ comoving Gauss coordinates.

In the general case, for asymptotically Minkowskian metrics, the divergence of $Q^{i00}$, $\partial_i Q^{i00}$, in asymptotically  rectilinear Minkowskian coordinates provides the integrand for the spatial volume integral giving the metric energy, $P^0$ (see Ref. \refcite{Weinberg}). If the metric is smooth enough, we can transform this 3-volume integral to a 2-surface integral over the infinite 3-space boundary, this common energy being the ADM mass.
Here, more specifically,  we are dealing with expression (\ref{Qi00}) in the marginally bound collapse ($k=0$) case, the volume integral giving the Weinberg metric energy $P^0$,  or ADM energy, 
\begin{equation} \label{Wenergy}
P^0= - \frac{1}{8 \pi}\int \partial_i Q^{i00} d^3 \rho 
\end{equation}
becoming
\begin{equation} \label{Wenergy-marginal}
P^0=   \frac{1}{8 \pi}\int \frac{\partial}{\partial \rho_i} \Big[ (A- \rho A')^2 \, \frac{n_i}{\rho^3} \Big] d\rho_1 d\rho_2 d\rho_3, 
\end{equation}
where we have taken $8 \pi$ for the Einstein constant.

We point out that, in the next subsection,  the expression (\ref{Wenergy-marginal}) for $P^0$, will be calculated in {\em intrinsic} coordinates, that is comoving Gauss coordinates, adapted to the spherical symmetry of the metric, and to its asymptotically Minkowskian character, and being {\em at rest} at the 3-space infinity with respect to the static Schwarzschild observers,  all this according to what have been explained in Sec. \ref{sec:2}. Because of these {\em intrinsic} coordinate properties, according to Refs. \refcite{Lapiedra-Morales-2013,LaMo-2014}, the resulting energy $P^0$ will be called the {\em intrinsic} energy of the corresponding metric (for some additional comments about its physical meaning see Sec. \ref{sec:7}). We make the natural assumption  that this {\em intrinsic} energy is the one which has to vanish in order that we can raise the question whether the corresponding metric could be quantically creatable from ``nothing'' (see for instance Refs. \refcite{Vilenkin-1983,Vilenkin-1999}).
%

%%%%%%%%%%%%%
%           %
%  Subsection 6A  %
%           %
%%%%%%%%%%%%%

\subsection{Calculating the {\em intrinsic} energy of the truncated matching metric}
\label{sec:6A}

Then, let us calculate this {\em intrinsic} energy $P^0$ of the above  truncated matching metric. Since now $Q^{i00}$ is regular enough (its first derivatives are continuous everywhere, except for $\rho = 0$) we can apply the Gauss theorem to the 3-volume in (\ref{Wenergy-marginal}) and express it as a 2-surface integral on the boundary. More specifically this boundary will be made of two 2-surfaces, $\rho = + \infty$ and $\rho = \epsilon >0$ where $\epsilon$ is a positive infinitesimal quantity. Then, we will take the limit $\epsilon \to 0$. 

So, in an evident notation we will have
\begin{equation} \label{P0-dos-sumands}
P^0=  P^0_{\infty} + \lim_{\epsilon \to 0} P^0_\epsilon, 
\end{equation}
with
\begin{equation} \label{P0-infty}
P^0_{\infty} = \lim_{\rho \to + \infty} \frac{1}{8 \pi} \int_{S_\rho} Q \, \cos \theta d \theta d \phi = \frac{1}{2}  \lim_{\rho \to + \infty} Q
\end{equation}
where the double integral is calculated on the 2-sphere of radius $\rho$, $S_\rho$, and
\begin{equation} \label{P0-epsilon}
P^0_{\epsilon} = - \frac{1}{2}  Q|_{\rho = \epsilon}, 
\end{equation}
and where
\begin{equation} \label{Q}
Q \equiv \frac{1}{\rho} (A - \rho A')^2.
\end{equation}

To calculate easily the limit (\ref{P0-infty}), notice that for $\rho > \lambda$ (and so for $\rho \to \infty$) our truncated metric is the Schwarzschild metric (\ref{LemaitreP}), 
that is, (\ref{LTB0}) with (\ref{k0}) and (\ref{Mrho}). Then an easy calculation gives for $\rho > \lambda$, $q$ having being defined at Eq. (\ref{qtauc}), 
\begin{equation} \label{Q-Sch}
Q = \displaystyle{q^2 \,  \frac{\tau^2}{\rho \Big(\tau - \frac{2}{3} \frac{\rho^{3/2}}{\sqrt{2m}}\Big)^{2/3}}},
\end{equation}
whose limit for $\rho \to \infty$ , and any attainable $\tau$ value, vanishes. Notice that we cannot put there $\tau \geq \psi (\rho = \lambda)$, since for this value of $\rho$ the outer spherical shell of the star has just reached its own singularity and we have no more a classical object ruled by General Relativity. According to (\ref{psi-simple}) 
this upper classical limit of $\tau$ is  $\tau =  \frac{2}{3} \frac{\lambda^{3/2}}{\sqrt{2m}} \equiv \tau_c$ (see Eq. (\ref{qtauc})). 

In all, the contribution $P^0_\infty$ to the total $P^0$ vanishes and we are left with the other contribution $\lim_{\epsilon \to 0} P^0_\epsilon$. Let us calculate it. First, according to (\ref{A-prima}) with $\varepsilon = -1$, we can write $Q$ as
\begin{equation} \label{Q-dev}
Q = \rho \Big(\frac{9M}{2(\tau - \psi)}\Big)^{2/3} \Big[\Big(\frac{1}{\rho} 
- \frac{1}{3}\frac{M'}{M}\Big) (\tau - \psi)+ \frac{2}{3} \psi' \Big]^2.
\end{equation}

Then, we are going to calculate $P^0$ for $\tau < \psi(0)$ since, as already commented, for $\tau = \psi(0)$ the inner spherical shell of the star reaches the intrinsic singularity and full General Relativity is no more valid. But, according to (\ref{psi-simple}), $\psi(\rho)$ is a growing function  of $\rho$, which means that in (\ref{Q-dev}) $\tau < \psi(\rho)$, $\forall \rho$. Thus, in order to calculate $\lim_{\rho \to 0} Q$, we only have to study how  the functions $M$ and $M'$, present in (\ref{Q-dev}), behave in this limit. To begin with, let us assume that in (\ref{Mrho-simple}) we have $\alpha \neq m$. In such a case, having in mind (\ref{Mrho-simple}) for $\rho < \lambda$, the limiting behaviour of $M$ and $M'$ when $\rho \to 0$ is
\begin{equation} \label{MM-prima-origen}
M \to m - \alpha \neq 0,  \qquad M' \to \frac{6 \alpha}{\lambda^2} \, \rho.
\end{equation}

From this we can easily see that there is a unique term in $Q$, Eq. (\ref{Q-dev}), the first one,  which diverges as $1/\rho$
when $\rho \to 0$. All in all, in (\ref{P0-dos-sumands}), the first summand,  $P^0_{\infty}$, vanishes, while the limit when $\epsilon \to 0$ of the second one, $P^0_{\epsilon}$, of our ``truncated matched metric'' when $\alpha \neq m$ diverges. In all, $P^0$ diverges, which makes implausible the existence in Nature of the corresponding metric: have in mind that this metric corresponds initially to a regular finite  physical system, more specifically to a finite self gravitating mass before reaching the last collapse stages, a gravitating mass that by no means should exhibit an infinite {\em intrinsic} energy.
\begin{figure}
\centerline{
\parbox[c]{0.7\textwidth}{\includegraphics[width=0.6\textwidth]{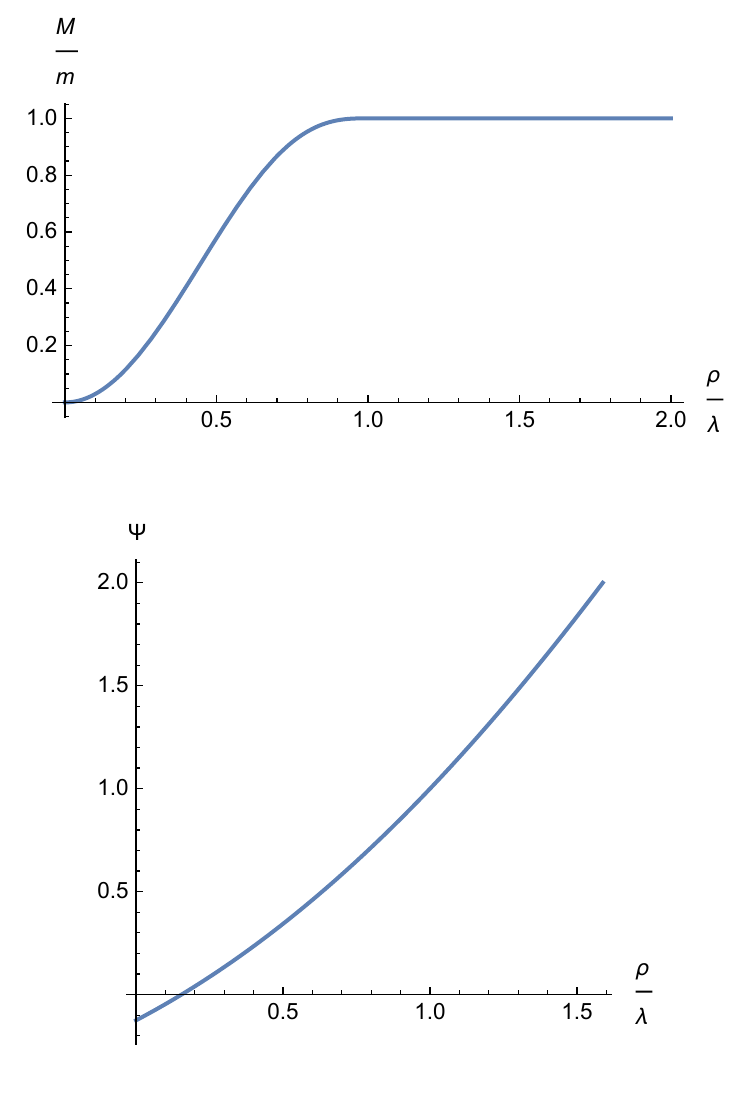}}} 
\caption{The functions $M(\rho)$ (top) and $\psi(\rho)$ (bottom) in the case of the simple ($\alpha = m$) matching metric for a marginally bound dust collapse model. Their first and second derivatives  are continuous across the matching surface  $\{\Sigma: \rho = \lambda$\}, ensuring the natural fulfillment  of the Lichnerowicz matching conditions  across $\Sigma$. Normalized variable, $\frac{\rho}{\lambda}$, and normalized functions, $\frac{M}{m}$ and $\Psi \equiv \frac{3}{2}\frac{\sqrt{2m}}{\lambda^{3/2}} \psi$, are used. \label{simple-collapse}}\end{figure}

Furthermore, in the parlance of Ref. \refcite{Lapiedra-Morales-2013}, the  ``truncated matched metric'' ($\alpha \neq m$) would  be non ``creatable'' 
because of the non vanishing value of its {\em intrinsic} energy.

Contrarily to this, following the precedent reasoning  in the present section, it is straightforward to see that the ``truncated matched metric'' with $\alpha = m$, that we will call henceforth  the ``simple matching metric'', has a vanishing {\em intrinsic} energy, $P^0$, and so a finite and constant one, that by itself would not be incompatible with the existence in Nature of the $\alpha = m$ metric (differently to the case $\alpha \neq m$ with, as we have seen, a divergent $P^0$). Then, because of this vanishing, it could be a metric candidate to be quantically creatable from ``nothing''. \cite{Vilenkin-1983,Vilenkin-1999}

The functions $M(\rho)$ and $\psi(\rho)$ for the ``simple matching metric'' with $\alpha = m$ are represented in Figure \ref{simple-collapse}. We can see that these two functions,  $\tau(\rho)$ and $\psi(\rho)$, are both non-decreasing functions, which according to (\ref{A-prima-zero}) leads to inequality $\tau(\rho) > \psi(\rho)$, $\forall \rho$, that is, in this case, the vanishing of $A'$ would happen after the time singularity which is non sense. Notice that this is a particular case of what was already mentioned in the paragraph following Eq. (\ref{sin-ese}).

For completeness we give here the mass density function $\mu(\tau, \rho)$ for this ``simple matching metric''. First, let us introduce the dimensionless variables
\begin{equation}\label{dimensionless}
t \equiv \frac{\tau}{m},  \qquad  \qquad x \equiv \frac{\rho}{\lambda}, 
\end{equation}
and the dimensionless parameter%
%%%FOOTNOTE-6
%
\footnote{See Ref. \refcite{LaMo-2017} for the role that this parameter plays in connection with a closely related collapse situation. \label{xi-parameter}}
%
%%%%END- FOOTNOTE-6
% 
\begin{equation}\label{xi-parameter}
\xi = \frac{\lambda}{2m}.
\end{equation}

Substituting  (\ref{Mrho-simple}) and (\ref{psi-simple}) for $\rho < \lambda$ and $\alpha = m$ in  (\ref{mu-dust}), the dust density becomes in this case:
\begin{equation}\label{mu-dust-simple}
\mu (t, x) = \frac{2}{\pi m^2} \, \frac{(x^2 -1)(x-1)} {\big[\xi^{3/2} p(x) -2t\big] \Big[\big[\xi^{3/2} p(x) -2t\big] 3 (x^2-1)(x-1) +  2 \xi^{3/2} x (x^4-3x^2 +3) \Big]}
\end{equation}
where, for convenience, we have denoted 
\begin{equation}\label{p-polonomi}
p(x) \equiv x^2 + 2x -\frac{1}{3}.
\end{equation}

\begin{figure}
\centerline{
\parbox[c]{0.9\textwidth}{\includegraphics[width=0.8\textwidth]{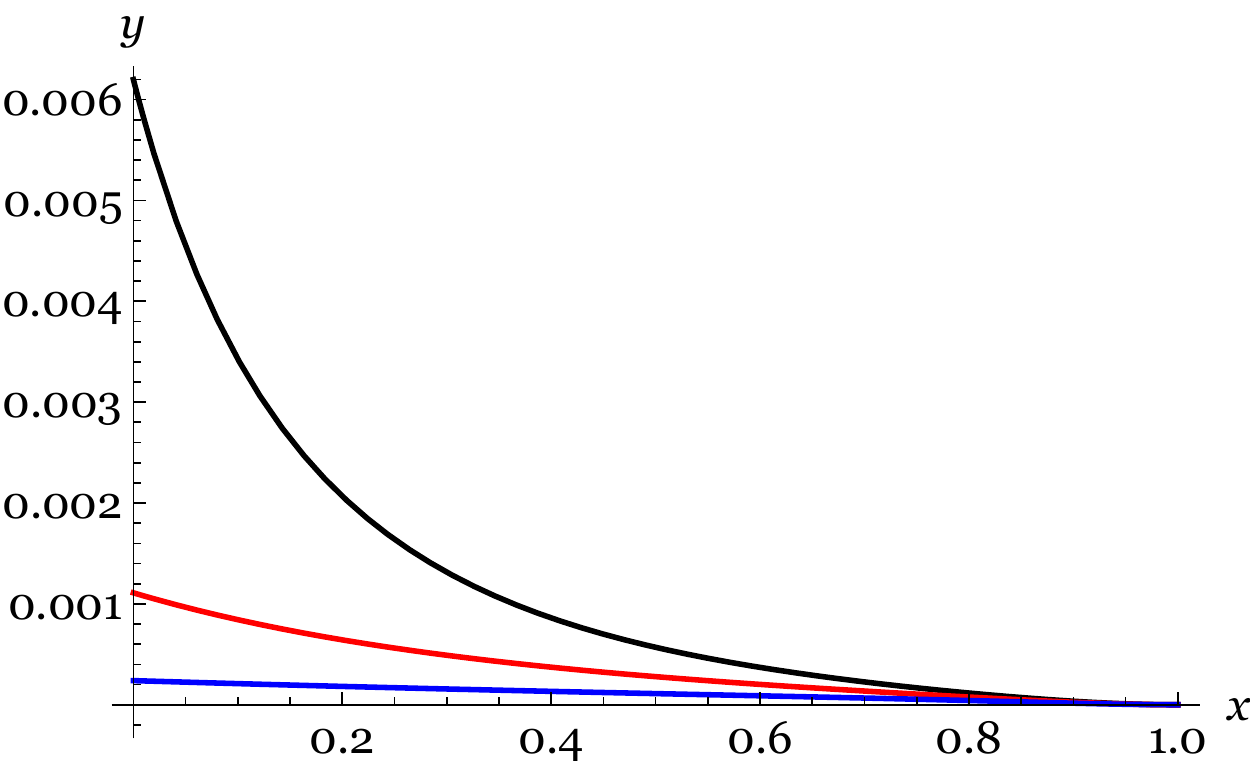}}} 
\caption{ Density profiles of the simple matching metric for $\xi =4$. The upper (black), middle (red) and lower (blue) lines,  correspond to $t=-5$, $t=-10$ and $t=-20$,  value of $t$ time, respectively.\label{Perfils-densitat}}\end{figure}

Using {\em Mathematica} software, it is easy to study the behaviour of the mass density function given by Eq. (\ref{mu-dust-simple}),   as well as the density profiles curves given by $\mu(\tau_b, \rho)$ at any fixed time before the essential singularity time of the $\rho$-shell,  $\tau_b < \psi(\rho)$. This have been carried out for $\xi = 2, 20, 200, 2000, ...$. All the graphics show that, for any $\tau_b$, the density profile is finite at the center, monotony decreasing for increasing $\rho \in [0, \lambda]$,  and such that 
\begin{equation}\label{limit}
\lim_{\rho \to \lambda} \mu(\tau_b, \rho) = 0 \qquad \forall \tau_b < \psi(\rho). 
\end{equation}
See, in particular, the specific case represented in Fig, \ref{Perfils-densitat},  where $y \equiv \pi m^2 \mu /2$.

Furthermore, it is easy to verify, following a similar way, that the generic matching metric has also a vanishing {\em intrinsic} energy, $P^0$, provided  that we fit the coefficients $M_k$ in (\ref{Mrho2}) such that the resulting $M(\rho)$ series expansion begins at least with the quadratic power of  $\rho$.%

Notice that there is no contradiction between the above $P^0$ vanishing and the well known result $P^0 = m$ for any spherical metric smoothly matched  to the Schwarzschild outer metric in static coordinates, 
since we use here {\em intrinsic} coordinates (see Sec. \ref{sec:7} for further discussions).

Now, a few words about the  uniqueness of the defined {\em intrinsic} coordinates, in relation to the freedom gauge, $\rho \to \bar{\rho}(\rho)$,   mentioned at the penultimate paragraph of Sec. \ref{sec:2}. Such freedom should be constrained such that, fast enough, 
 $\bar{\rho} \to \infty$ when $\rho \to \infty$ and  $\bar{\rho} \to 0$ when $\rho \to 0$, in order to preserve the finite and stationary character of the {\em intrinsic} energy in the new coordinate  $\bar{\rho}$. Further, the function  $\bar{\rho}(\rho)$ should be ${\cal C}^1$ class in order to preserve the matching Lichnerowicz conditions.

%%%%%%%%%%%%%
%           %
%  Subsection 6B  %
%           %
%%%%%%%%%%%%%

\subsection{Calculating the {\em intrinsic} energy of the Oppenheimer-Snyder metric}
\label{sec:6B}

To finish the present section, let us now show that the Oppenheimer-Snyder (O-S) metric studied in Sec. \ref{sec:4} has a non finite and not constant, and so a non vanishing value for its {\em intrinsic} energy, differently to our ``simple matching metric'' ($\alpha = m$). The O-S metric would be, then, a non quantum ``creatable'' metric. 
We point out here this creativiness question  only to remark how great can be the difference between metrics that do not fulfill the same matching conditions, despite the fact that all them fulfill the same evolving differential equations and share very similar sources. 

The first thing to be noticed in order to calculate $P^0$, in the O-S case, is that now the integrand in the 3-volume integral of (\ref{Wenergy}) is no more regular every where (besides the intrinsic singularity at $\tau = \psi(\rho)$), since as already remarked this metric does not match continously through the 
3-surface $\rho=\lambda$. More specifically, the function $A'$ is discontinuous at $\rho = \lambda$ according to (\ref{A-prima-OS-dis}).

Then, similarly to what has been done in the case of the ``simple matching metric'',   we will cast the 3-volume integral in  (\ref{Wenergy})  to a 2-surface integral on the appropriate boundary, although now this boundary is made out of four 2-surfaces: the $\rho \to \infty$ 
and the $\rho \to 0$, plus two more ones, $\rho = \lambda + \epsilon$ and  $\rho = \lambda - \epsilon$, with $\epsilon \to 0$  ($\epsilon >0$). Therefore $P^0$ will be the sum of four different limit contributions, from four 2-surface integrals, that in an evident notation can be written as $P^0_\infty$,  $P^0_\epsilon$, $P^0_{\lambda +\epsilon}$, 
$P^0_{\lambda -\epsilon}$. 

The first contribution $P^0_\infty$ vanishes again, since for $\rho > \lambda$ (and then for $\rho \to \infty$) the O-S metric becomes the exterior Schwarzschild metric, like in the precedent case of  the ``simple matching metric'', and so, like in this case, we have  $P^0_\infty = 0$. 

The second contribution, $P^0_\epsilon$, vanishes too, since according to (\ref{A-function-OS}), for $\rho < \lambda$ (and so for $\rho = \epsilon$ too), the corresponding $A$ function is linear in $\rho$, which means that $Q$ in (\ref{Q}) vanishes identically, except for 
$\tau = \frac{2}{3} \frac{\lambda^{3/2}}{\sqrt{2m}}\equiv \tau_c$. But, for the O-S metric, $\tau_c$ is the time of the simultaneous collapse, such that we only have to consider times $\tau < \tau_c$ if we want to remain in the classical frame of General Relativity.

Then, we are left with the remaining contributions coming from $P^0_{\lambda +\epsilon}$ and  $P^0_{\lambda -\epsilon}$, that is
\begin{equation} \label{P0-OS}
P^0 =  \lim_{\epsilon \to 0} (P^0_{\lambda + \epsilon} + P^0_{\lambda - \epsilon}) =  \frac{1}{2}\lim_{\epsilon \to 0}(Q|_{\lambda+\epsilon} -Q|_{\lambda-\epsilon}) 
\equiv \frac{1}{2} [Q]_{_{\Sigma}}
\end{equation}

Having in mind (\ref{Q})  and using the fact that $A$ is continuous through $\rho = \lambda$, we obtain 
\begin{equation} \label{P0-OS-2}
P^0 =  \Big(\frac{\lambda}{2} \,  [A'^2]_{_{\Sigma}} - A(\tau, \rho = \lambda)  [A']_{_{\Sigma}}\Big), 
\end{equation}
and using (\ref{A-prima-OS-dis}) and (\ref{A-prima-quadrat-OS-dis}) one finally obtains
\begin{equation} \label{P0-OS-3}
P^0 = \frac{q^2}{2 \lambda} \, \frac{\tau^2}{(\tau  - \tau_c)^{2/3}}, 
\end{equation}
for $\tau < \tau_c$, as we have explained before.

Thus, for the O-S metric, the {\em intrinsic} energy $P^0$ does not vanish. Further, it depends on $\tau$ and grows without limit when $\tau$ approaches $\tau_c$. These last properties do not seem good news for the physical character of the O-S metric, since this is the asymptotically Minkowskian metric of a finite system ($m$ and $\lambda$ are finite) which remains isolated, neither absorbing, nor emitting, gravitational or electromagnetic radiation, and further without accreting or expelling any matter. Actually, because of these characteristics, the {\em intrinsic} $P^0$ value for the O-S metric would be expected to be finite and to remain constant. This is just what happens to our ``simple matching metric'' for instance, its {\em intrinsic} $P^0$ being zero. In this way, the physical character of this last metric would be enhanced.
Therefore, and according to our point of view,  the O-S metric would not be a good metric candidate to actually exist in Nature, but the ``simple matching metric''  could in principle be such a candidate. 

Whatever could it be, and in another words, in the present paper, we have assumed both, the Lichnerowicz matching conditions and  the existence of a finite stationary intrinsic energy, and we have explored the consequences of these two assumptions on the corresponding metrics of our LTB metric family. Obviously, the goodness of the metric solutions found in this way rely on the goodness of these, in any case, interesting assumptions.

%%%%%%%%%%%%%
%           %
%  SECTION  7  %
%           %
%%%%%%%%%%%%%

\section{Discussion}
\label{sec:7}

In the book by Hawking and Ellis \cite{Hawking-Ellis}, p. 58, the authors  assume that a suitable space-time would be represented by a ${\cal C}^2$ class differentiable manifold in accordance with a common point of view on the subject (see details in Ref. \refcite{Lichnerowicz}). But at the same time, those authors reduce the importance of the assumption made by making the following comment ``In fact, the order of differentiability of the metric is probably not physically significant. Since one can never measure the metric exactly, ...''. %

However, the results of our calculations in the present paper show that the consequences of imposing the Lichnerowicz matching conditions are far of having minor consequences. On the contrary, imposing them we have passed from the known O-S metric, that does not satisfy these conditions, to another metric, ``the simple matching metric'',  sensibly different either locally or globally. Thus, for instance,  this ``simple matching metric'' has a vanishing {\em intrinsic} energy, $P^0$, while $P^0$ for the O-S metric does not vanish, depending on time $\tau$ and growing up indefinitely when $\tau$ approaches the essential singularity. As commented at the end of the previous section, and letting aside the rough assumption of an stellar collapse with an everywhere vanishing pressure, this could make the existence in Nature of the O-S metric rather implausible. However, this possible difficulty would not be present in the case of the `simple matching metric" since it has a finite and constant intrinsic energy (more specifically, a vanishing one). 

The concept of the energy $P^0$ of a metric, as derived from an energy momentum complex (the Weinberg one in our case), is not always considered a physical quantity, but is always considered as such for asymptotic Minkowskian metrics, in the appropriated coordinates. All the metrics studied in the present paper are asymptotic Minkowskian metrics, in particular the ``simple matching metric''  and the O-S ones. So, in our case, $P^0$, although coordinate dependent (as it could not otherwise be), should be considered in principle as some kind of physical energy in some precise sense. 
As it is well known, for asymptotic Minkowskian metrics,  $P^0$, when calculated in asymptotic Minkowskian coordinates, is the time component of a Minkowskian linear 4-momentum vector. However, what we have calculated here is the {\em intrinsic} value of $P^0$ which, by definition, is not calculated in merely asymptotic Minkowskian coordinates, but in {\em intrinsic} coordinates: the spherical symmetrical coordinates associated to a congruence of everywhere free falling observers, comoving with the infalling pressureless matter, asymptotically Minkowskian and at rest at the 3-space infinity. As a result, the {\em intrinsic} value of $P^0$ is no more the time component of a Minkowski 4-vector. But this does not change the fact that this {\em intrinsic} energy remains a total conserved quantity whose different parts, the relativistic kinetic energy of the matter and the gravity contribution, enter in a mutual balance. This kind of balance is constitutive of the definition of total energy, either in the present case or in another similar one of the rest of classical physics. Here the balance is ensured by the conservation of the (Weinberg) energy-momentum complex, that is by the vanishing of its ordinary divergence.
In all, our {\em intrinsic} energy $P^0$ is a physical energy, in the above {\em intrinsic} coordinates, that as such should remain finite and constant for any metric whose source be an isolated, finite, regular, mass distribution, as it is initially the kind of mass distributions considered in the present paper.
See again the penultimate paragraph of subsection \ref{sec:6B} in relation to this.

Some more comments on the physical meaning of the metric energy and {\em intrinsic} metric: in order that we can recover for the energy defined in this way the usual properties of the energy of a non-gravitational system, we should use appropriate coordinate systems among all the arbitrary systems of the general relativity. First, these appropriate coordinates should be ``realized with real bodies'', that is, ``the triple of space coordinates $x^1, x^2, x^3$, can be any quantities defining the position of bodies in space, and the time coordinate $x^0$ can be defined by an arbitrary running clock'' (see Landau and Lifshitz, The theory of classical fields, Pergamon Press, 1971, epigraph 84). Further, the resulting coordinates must be well adapted to the asymptotic Minkowskian character of the metric if such character were the case.

Even with these constraints, many different physically meaningful coordinate systems can exist, to each one corresponding a different value of the above defined metric energy, this difference in the values being in accordance with what already happens in the absence of any gravitational field. Can, then, Gauss coordinates (that is, coordinates in which $g_{00} = -1, g_{0i}=0$) be one of these meaningful coordinate systems? Yes, they can: to begin with they are "realized with real bodies", but besides it they are particularly suitable coordinates, the reason being the following one: the defined energy of the gravitational field and its source is, as a 3-volume integral, the addition of many (actually infinite in number) elementary energy contributions, all them taken in the same time coordinate. But in general, this time is not a ``synchronous time'' (see again the above epigraph 84), so that we can be adding non simultaneous contributions in order to calculate the total energy, which is not a convenient situation. Needless to say that this problem is no present in non-gravitational systems, but it is neither present in presence of gravitation if we take Gauss coordinates, since in such a case the time is synchronized. Thus, in Gauss coordinates we recover the behavior of the defined energy in non-gravitating systems, that is, that the total energy becomes the addition of its simultaneous elementary contributions.

Then, what about our intrinsic  energy? Is it a physical value of our defined energy? To begin with, it is calculated in Gauss coordinates, which as we just have seen are particularly suitable coordinates to produce a physical energy value. But in accordance with its denomination, {\em ``intrinsic''}, and trying to generalize what could be call {\em intrinsic} energy of a system of free particles in special relativity, we want an energy value which be due to the metric itself, without the mentioned ``fictitious'' partial contributions, and without other partial contributions raised by observers not belonging to our measured systems (like, in our case, static external observes, instead of observers tied to source matter freely falling dawn). This is why the particular Gauss coordinates used in the definition of our intrinsic energy are, in the case of the considered metrics, co-moving with the source matter, adapted to the asymptotically Minkowskian character, and to the spherical symmetry, and at rest at the 3-space infinity, with respect to ordinary static Schwarzschild coordinates.

The just above definition of {\em intrinsic} coordinates, for the simple metrics considered in the present paper, is a particular case of a larger, non-unique, {\em intrinsic} coordinate definition given in Ref. \refcite{Lapiedra-Morales-2012}.

Now, before finishing the present section, in according with what has been mentioned at the Introduction, it is known that the O-S metric provides important information when it is taken as a test-bed in numerical relativity. In our opinion, the well matched metrics in the present paper can play a similar role even better, since now all the initial conditions for the numerical integration of the Einstein equations will be continuous functions everywhere (out of the essential singularity, of course), and more specifically across the matching surface. In principle, such properties could avoid spurious oscillations due to the Gibbs phenomena (see, for instance, Ref. \refcite{Sod}) making the calculations more robust.

To finish this section, we find that our present work  allows us to select some solutions mathematically and physically significant from a large family of metric solutions of Einstein field equations, i.e., the pressureless  LTB solutions for a finite mass. The tentative selection criterium is double: fulfillment of smooth enough matching conditions, and asking for a finite, constant value of the corresponding {\em intrinsic} energy.

%\vspace{1cm}

%%%%%%%%%%%%%%%%%%
%           %
%    ACKNOWLEDGMENTS   %%
%           %
%%%%%%%%%%%%%%%%%%

%%%%%%%%%%%%%%%%%%%%%%%%%%%%%%%%%%%%%%%%%%

\section*{Acknowledgements}
This work has been  supported by the Spanish ``Ministerio de Econom\'{\i}a y Competitividad'' and the ``Fondo Europeo de Desarrollo Regional'' MINECO-FEDER Project No. FIS2015-64552-P. Useful discussions with Jos\'{e} A. Font, Jos\'{e} Mar\'{\i}a Ib\'a$\tilde{\rm n}$ez and Bartolom\'e Coll 
are gratefully recognized. 

%%%%%%%%%%%%%%%%%%%%%%%%%%%%%%%%%%%%

%%%%%%%%%%%%%
%           %
%  APPENDIX A %
%           %
%%%%%%%%%%%%%

\appendix

\section{Lichnerowicz matching conditions and dust collapse}
\label{ap-A}

For the subclass of the LTB family of dust metrics given by (\ref{LTB0}), the Lichnerowicz matching conditions guarantee that the metric coefficients $A$ and $A'$ are function of class ${\cal C}^1$, by imposing the nullity of the following Hadamard discontinuities across the matching 3-surface $\{\Sigma: \rho = \lambda\}$:
\begin{equation}\label{LichneR}
  [A]_{_\Sigma} = [A']_{_\Sigma} = [A'']_{_\Sigma} = 0.
\end{equation}
from which the new equalities $[\dot{A}]_{_\Sigma} = [\dot{A}']_{_\Sigma} = 0$ follow. 

This appendix is devoted to establish the following result: 

Let it be a LTB metric (\ref{LTB0}) describing a spherically symmetric dust collapse in parabolic evolution and let us consider the matching with the exterior metric (\ref{LemaitreP}).  The Lichnerowicz matching conditions (\ref{LichneR}) are satisfied on the matching surface $\{\Sigma: \rho = \lambda\}$ if, and only if,  the free functions $M(\rho)$ and $\psi(\rho)$ (providing the function $A(\tau,\rho)$ from (\ref{k0})) and their derivatives are subject to the following conditions:
\begin{equation}\label{Lichne1}
[M]_{_\Sigma} = [\psi]_{_\Sigma} = [M']_{_\Sigma} = [\psi']_{_\Sigma} = 0, 
\end{equation}
and 
\begin{equation}\label{Lichne2}
[M'']_{_\Sigma} = \frac{3}{2} \varepsilon \Big(\frac{2M}{A}\Big)^{3/2}_{|\Sigma} \,  [\psi'']_{_\Sigma}. 
\end{equation}
where the proportionality factor between the discontinuities of the second derivatives  in  (\ref{Lichne2}) is calculated on $\Sigma$,  by putting $M(\lambda) = m$ and 
$A(\tau, \lambda) = \sqrt[3]{\frac{9}{2} m [\tau - \psi(\lambda)]^2}$,  with 
$\psi(\lambda) = \frac{2}{3} \frac{\lambda^{3/2}}{\sqrt{2m}}$ (see Eq. (\ref{Mrho})).

In order to prove this result, let us write the first derivatives of the function $A$, 
\begin{equation}\label{Ap-Apunt}
\dot{A} = \varepsilon \sqrt{\frac{2M}{A}}, 
\end{equation}
\begin{equation}\label{Ap-Aprima}
A' = \frac{1}{3} \frac{M'}{M} \, A - \varepsilon \sqrt{\frac{2M}{A}} \, \psi',  
\end{equation}
and express the second derivatives in this convenient way:
\begin{equation}\label{Ap-Apunt-punt}
\ddot{A} = - \frac{M}{A^2}, 
\end{equation}
\begin{equation}\label{Ap-Apunt-prima}
\dot{A}' = \varepsilon \frac{2}{3} \frac{M'}{\sqrt{2M}} \frac{1}{\sqrt{A}} + \frac{M}{A^2}  \, \psi', 
\end{equation}
\begin{equation}\label{Ap-Aprima-prima}
A'' = \frac{1}{3} \Big(\frac{M''}{M} \, A - \frac{M'^2}{M^2} \, A + \frac{M'}{M}\, A'\Big)- \dot{A}' \, \psi' - \dot{A} \psi''.
\end{equation}
Then, from Eqs. (\ref{k0}) and (\ref{Ap-Apunt}) we have:
\begin{equation}\label{Ap-A-Adot-dis}
[A] = [\dot{A}] = 0 \, \iff [M] = [\psi] = 0, 
\end{equation}
where, from now on,  the Hadamard discontinuities  will be written without subscript.

On the other hand, from Eqs. (\ref{Ap-Aprima}) and (\ref{Ap-Apunt-prima}), taking into account  (\ref{Ap-Apunt}) and (\ref{Ap-A-Adot-dis}) the following relations must be satisfied:
\begin{equation}
[A'] =  \frac{1}{3} \frac{A}{M} [M'] - \varepsilon \sqrt{\frac{2M}{A}}[\psi'], 
\end{equation}
\begin{equation}
[\dot{A}'] = \varepsilon \frac{2}{3}  \frac{1}{\sqrt{2M A}} [M'] + \frac{M}{A^2} [\psi'].
\end{equation}
The homogeneous linear system $[A'] = 0$ and  $[\dot{A}'] = 0$ in the two unknown  $[M']$ and $[\psi']$ has  non vanishing determinant $1/A \neq 0$,  and then,  its sole solution is the  trivial one, $[M'] = [\psi'] =0$. Consequently:
\begin{equation}\label{Ap-Aprima-Adot-prima-dis}
[A'] = [\dot{A}'] = 0 \, \,  \iff \,  \, [M'] = [\psi'] =0.
\end{equation}
Now, from Eq. (\ref{Ap-Aprima-prima}), and taking into account  that  (\ref{Ap-A-Adot-dis}) and  (\ref{Ap-Aprima-Adot-prima-dis}) ensures that $M$ and $\psi$ 
are function of class ${\cal C}^1$ in the whole domain of the used Gauss coordinates, we obtain for the discontinuities of the second derivatives the relation: 
\begin{equation}\label{A-2primabis}
[A''] = \frac{1}{3}  \frac{A}{M}\, [M''] -  \varepsilon \sqrt{\frac{2M}{A}} [\psi''], 
\end{equation} 
and the condition given by Eq.  (\ref{Lichne2}) follows:
\begin{equation}\label{A-2primabis-bis}
[A''] = 0 \iff [M''] = \frac{3}{2} \varepsilon \Big(\frac{2M}{A}\Big)^{3/2}_{|\Sigma} \,  [\psi''], 
\end{equation}
where we must take $\varepsilon = 1$ or $\varepsilon = -1$ for a  expanding or collapsing scenario, respectively. Further, since (\ref{A-2primabis-bis}) must remain valid for any value of $\tau$ (present in the $A$ function), the only way that (\ref{A-2primabis-bis}) can be fulfilled is having both $[M''] = [\psi'']=0$.

By construction,  for the ``generic matching metric'' analyzed in Sec. \ref{sec:4}, one has conveniently taken $[M''] = [\psi''] =0$, ensuring directly from (\ref{A-2primabis}) that $[A''] = 0$.

Finally, since $[A] = 0$, it must be $[\ddot{A}] = 0$, which is consistent with (\ref{Ap-Apunt-punt}).

\vspace{1cm}

\noindent {\bf Note.-} It is to be remarked that the obtained non-vanishing value of the intrinsic energy, $P^0$, of the Oppenheimer-Snyder metric [see Eq.  (\ref{P0-OS-3})] is bound to the fact that in the used intrinsic coordinates this metric is not a solution of the Einstein field equations in the strict sense on the star 2-surface. Actually, after this paper was published, we noticed that this energy value would become zero, as could be expected, provided we complete this metric in order to be everywhere a solution of the Einstein field equations, in the sense of the distributions (the calculation details will be published elsewhere).  However this vanishing result does not remain stable when we break slightly the original spherical symmetry of the problem. Then, the conclusions we draw in the published paper from the above non-vanishing of $P^0$ should perhaps be revised having in mind these last results.

%%%%%%%%%%%%%%%%%%
%           %
%    REFERENCES  %%
%           %
%%%%%%%%%%%%%%%%%%

%

%%%%%%%%%%%%%%%
%%%%%%%%%%%%%%%
\end{document}